\documentclass[twocolumn,english,aps,prd,nofootinbib,hidelinks]{revtex4-2}
\usepackage{lmodern}
\usepackage[LGR,T1]{fontenc}
\usepackage[latin9]{inputenc}
\usepackage{xcolor}
\usepackage{babel}
\usepackage{amsmath}
\usepackage{amssymb}
\usepackage{graphicx}
\usepackage[unicode=true]
 {hyperref}

\makeatletter

\DeclareRobustCommand{\greektext}{%
  \fontencoding{LGR}\selectfont\def\encodingdefault{LGR}}
\DeclareRobustCommand{\textgreek}[1]{\leavevmode{\greektext #1}}
\ProvideTextCommand{\~}{LGR}[1]{\char126#1}

\usepackage{enumitem}

\makeatother

\begin{document}
\title{New exact solution and $\mathcal{O}\left(1/\sqrt{\omega}\right)$
anomaly in Brans-Dicke gravity with\linebreak trace-carrying matter}
\author{Hoang Ky Nguyen$\,$}
\email[\ \ ]{hoang.nguyen@ubbcluj.ro}

\affiliation{Department of Physics, Babe\c{s}-Bolyai University, Cluj-Napoca 400084,
Romania}
\author{Bertrand Chauvineau$\,$}
\email[\ \ ]{bertrand.chauvineau@oca.eu}

\affiliation{Universit\'e C\^ote d'Azur, Laboratoire Lagrange (UMR 7293), CNRS,
Observatoire de la C\^ote d\textquoteright Azur, BP 4229, 06304 Nice
cedex 4, France}
\date{September 24, 2024}
\begin{abstract}
\noindent \vskip2ptWe present an exact static spherically symmetric
solution for the Brans-Dicke action sourced by a self-gravitating
massless Klein-Gordon helicity--0 field, which notably possesses
an energy-momentum tensor with \emph{non-vanishing trace}. Through
a Weyl mapping into the Einstein frame, the Brans-Dicke scalar field
transforms into a ``dilaton'' that couples with the Klein-Gordon field.
Despite this dilatonic coupling, the field equations of the resulting
Einstein-Klein-Gordon-dilaton action are fully soluble using the harmonic
radial coordinate. Remarkably, in the limit of $\omega\rightarrow\infty$,
the Brans-Dicke scalar field exhibits an anomalous behavior of $\mathcal{O}(1/\sqrt{\omega})$
as opposed to the expected $\mathcal{O}\left(1/\omega\right)$. Consequently,
the solution converges to a spacetime configuration of General Relativity
sourced by the original Klein-Gordon field and a free scalar field,
the latter being the ${\cal O}\left(1/\sqrt{\omega}\right)$ ``remnant''
of the Brans-Dicke scalar field. Although this $\mathcal{O}\left(1/\sqrt{\omega}\right)$
anomaly was previously identified for Brans-Dicke vacuum and Brans-Dicke-Maxwell
electrovacuum, our findings demonstrate its generic applicability
in Brans-Dicke gravity, \emph{irrespective} of the trace of the energy-momentum
tensor of the source. We also provide a formal mathematical proof
to bolster this conclusion. Furthermore, we discuss the potential
implications of the $\mathcal{O}\left(1/\sqrt{\omega}\right)$ anomaly
in improving the relativistic corrections to Newtonian gravity beyond
the weak-field parametrized post-Newtonian formalism.
\end{abstract}
\maketitle

\section{\label{sec:Motivation}Motivation}

The late-time cosmic accelerated expansion observed in 1998 has ignited
a keen interest in modifying General Relativity (GR) to accommodate
the enigmatic ``dark energy'' component \citep{reviews-MG}. Simultaneously,
early-time cosmic inflation scenarios often necessitate extensions
beyond GR, for example, the Starobinsky model ${\cal R}+{\cal R}^{2}$
\citep{Starobinsky-1980}. A very active avenue of modified gravity
involves introducing a scalar degree of freedom alongside the well-established
tensor components in GR \citep{reviews-ST,Fujii-2003}. Scalar degrees
of freedom naturally emerge in various theoretical frameworks, including
Kaluza-Klein, string theory, and braneworld scenarios \citep{Fujii-2003}.\vskip4pt

A broad class of scalar-tensor theories is encapsulated by the Bergmann-Wagoner
action \citep{Bergmann-1968,Wagoner-1970}
\begin{equation}
S_{\text{ST}}=\int d^{4}x\frac{\sqrt{-g}}{16\pi}\left[\phi\mathcal{R}-\frac{\omega(\phi)}{\phi}\left(\nabla\phi\right)^{2}-2V(\phi)\right]+S^{(m)}\label{eq:BW-action}
\end{equation}
This class encompasses $f({\cal R})$ theories (when $\omega(\phi)\equiv0$,
upon certain transformations to re-express the $\phi$ field in terms
of ${\cal R}$) and the pure Brans-Dicke (BD) theory ($\omega(\phi)\equiv\omega$,
where $\omega$ is a constant, and additionally, $V(\phi)\equiv0$).
The interplay of the scalar field and the metric tensor is a non-trivial
aspect of these theories.\vskip4pt 

Of crucial importance is the limit of the scalar-tensor action \eqref{eq:BW-action}
to GR. A necessary---but not sufficient---condition for a recovery
of GR is that $\phi$ becomes a positive-definite \emph{constant}
field, viz. $\phi\equiv\bar{\phi}>0$, in which case $8\pi/\bar{\phi}$
plays the role of a Newtonian gravitational constant and $V(\bar{\phi})/\bar{\phi}$
acts like a ``cosmological constant''. A common way to drive the field
$\phi$ to a constant value is by allowing $\omega(\phi)$ to tend
to infinity, for example, via the Damour-Nordtvedt dynamical attractor
mechanism \citep{Damour-1993}. For the simplest scalar-tensor version---BD
gravity---in which case the Damour-Nordtvedt mechanism is not applicable,
a constant field $\phi$ would be achieved by sending the BD parameter
$\omega$ to infinity \citep{Weinberg}.\vskip4pt

The large--$\omega$ limit is subtle and delicate, at least for the
pure BD version, however. Although the variation $\left(\nabla\phi\right)^{2}$
dies off as $\omega$ grows to infinity, their combination $\omega\left(\nabla\phi\right)^{2}$
\emph{may} remain significant. There is no \emph{a priori} theoretical
reason that the ``kinetic'' term $\omega\left(\nabla\phi/\phi\right)^{2}$
in action \eqref{eq:BW-action} must be negligible compared with the
Einstein-Hilbert term ${\cal R}$ upon the infinite $\omega$ limit.
From the variational principle standpoint, when the action expressed
in \eqref{eq:BW-action} is extremized, the two said terms are in
competition, and in principle, they can attain \emph{comparable magnitudes}.\vskip4pt

This scenario is precisely what happens in BD gravity, revealing a
non-convergence behavior that has been extensively explored and documented
since the 1990s. Specifically, various authors have established that
the infinite $\omega$ limit fails to transition the BD vacuum to
a vacuum of GR \citep{Romero-1993-a,Romero-1993-b,Romero-1993-c,Romero-1998,Romero-2021,Faraoni-1998,Faraoni-1999,Faraoni-2019,Bhadra-2001,Bhadra-2002,Bhadra-2005,Fabris-2019,Quiros-1999,Banerjee-1997,Faraoni-2018}.
Instead, it approaches a GR solution sourced by a free massless scalar
field; (note: for static spherically symmetric setup, the limiting
vacuum is the Fisher-Janis-Newman-Winicour (FJNW) solution \footnote{The solution has been re-discovered several times, and is also known
as the Fisher-Bergmann-Leipnik-Janis-Newman-Winicour-Buchdahl-Wyman
(FBLJNWBW) solution \citep{Faraoni-2018}.}). Moreover, the convergence rate in the limit is determined to be
${\cal O}\left(1/\sqrt{\omega}\right)$ instead of ${\cal O}\left(1/\omega\right)$.
That is to say, the BD scalar field behaves as
\begin{equation}
\phi\simeq\bar{\phi}\,\biggl[1+\frac{\stackrel{1}{\varphi}}{\sqrt{\omega}}+{\cal O}\left(\frac{1}{\omega}\right)\biggr]\label{eq:remnant-1}
\end{equation}
with $\bar{\phi}$ being a non-zero constant and the sub-leading field
$\stackrel{1}{\varphi}$ \emph{non-constant}. We shall aptly call
$\stackrel{1}{\varphi}$ a ``remnant'' BD field in the infinite $\omega$
limit.\vskip4pt

These findings are two sides of the same coin. Due to the $\stackrel{1}{\varphi}$
associated with the $1/\sqrt{\omega}$ order, the ``kinetic'' term
remains \emph{finite}, viz.
\begin{equation}
\omega\left(\nabla\phi/\phi\right)^{2}\simeq\bigl(\ \nabla\stackrel{1}{\varphi}\:\bigr)^{2}\label{eq:remnant-2}
\end{equation}
and contributes to the limiting vacuum, causing deviation from the
GR vacuum. The role of the ``remnant'' field $\stackrel{1}{\varphi}$
is crucial: it acts as an (additional) free massless scalar field
sourcing the FJNW solution mentioned above. \vskip4pt

This $\mathcal{O}\left(1/\sqrt{\omega}\right)$ \emph{anomaly}, as
shall be termed in our present article, has also been confirmed for
electrovacuum, namely, the BD action coupled with electromagnetism,
known as the Brans-Dicke-Maxwell theory \citep{Faraoni-1998,Faraoni-1999,Faraoni-2019,Bhadra-2002,Bhadra-2001,Bhadra-2005}.
As the energy-momentum tensor (EMT) of the Maxwell electromagnetic
field is traceless, the ${\cal O}\left(1/\sqrt{\omega}\right)$ anomaly
has been asserted to exist for BD gravity sourced by \emph{traceless}
matter (a condition trivially met in pure vacuum).\vskip4pt

The question concerning fields with a \emph{non-vanishing} EMT trace
remains in contention, however \citep{Faraoni-1998,Faraoni-1999}
\footnote{Note: it has been advocated in \citep{Faraoni-1998,Faraoni-1999}
that the ${\cal O}\left(1/\sqrt{\omega}\right)$ anomaly should be
confined to traceless matter as this type of matter possesses certain
conformal symmetry, while trace-carrying matter lacks such symmetry.
This line of reasoning is \emph{circumstantial}, however, as the ${\cal O}\left(1/\sqrt{\omega}\right)$
anomaly might stem from a cause independent from the conformal symmetry
at play.}. Normal matter, that comprises stars, in general possesses non-vanishing
EMT trace, such as a non-relativistic perfect fluid with an EMT given
by $T_{\mu}^{\nu}=\text{diag}\left(-\rho,P,P,P\right)$ where $P\ll\rho$.
Consequently, \emph{if} the ${\cal O}\left(1/\sqrt{\omega}\right)$
anomaly is proven to hold for trace-carrying matter, it could have
implications in observational astronomy \citep{Chauvineau-2007}.
For instance, it may indicate the need to \emph{go beyond} the Robertson
parameter $\gamma$ derived in the weak-field parametrized post-Newtonian
(PPN) approximation, given by $\gamma_{\text{PPN}}=\frac{\omega+1}{\omega+2}$,
\emph{which lacks the ${\cal O}\left(1/\sqrt{\omega}\right)$ hallmark.}\vskip8pt

We plan to tackle this problem in a two-pronged approach. Firstly,
we shall present a mathematical proof of the ${\cal O}\left(1/\sqrt{\omega}\right)$
anomaly in BD gravity coupled with matter fields of \emph{non-vanishing}
EMT trace. This proof builds upon preliminary ideas previously ventured
by one of us in \citep{Chauvineau-2003,Chauvineau-2007}. We find
that the ``remnant'' field $\stackrel{1}{\varphi}$ continues to function
as a free massless scalar field in the infinite $\omega$ limit. Its
contribution to the ``kinetic'' term, as per Eq. \eqref{eq:remnant-2},
and the $1/\sqrt{\omega}$ factor associated with the ``remnant''
field $\stackrel{1}{\varphi}$ via Eq. \eqref{eq:remnant-1} persist
independently of the EMT trace of the matter source.\vskip4pt

Secondly, we shall reinforce this proof by examining a concrete case:
the BD theory sourced by a massless Klein-Gordon helicity-0 field.
For the static spherically symmetric setup, this theory proves to
be \emph{soluble}, and the $\mathcal{O}\left(1/\sqrt{\omega}\right)$
signature manifests in the final analytical solution. As a by-product,
the exact solution obtained herein can also serve as a prototype for
future studies involving the BD action coupled with trace-carrying
matter fields. \vskip4pt

A massless Klein-Gordon scalar field $\Psi$ is characterized by an
EMT given by
\begin{equation}
T_{\mu\nu}^{\Psi}=\partial_{\mu}\Psi\partial_{\nu}\Psi-\frac{1}{2}g_{\mu\nu}g^{\lambda\rho}\partial_{\lambda}\Psi\partial_{\rho}\Psi
\end{equation}
carrying a non-zero trace, $T^{\Psi}=-g^{\mu\nu}\partial_{\mu}\Psi\partial_{\nu}\Psi$.
In contrast, the Maxwell electromagnetic (EM) field has a traceless
EMT 
\begin{equation}
T_{\mu\nu}^{F}=-F_{\mu\lambda}F_{\nu}^{\ \lambda}+\frac{1}{4}g_{\mu\nu}F_{\lambda\rho}F^{\lambda\rho}\ \ \Longrightarrow\ \ T^{F}=0
\end{equation}
The Brans-Dicke-Maxwell (electro)vacuum was first obtained by Bronnikov
\citep{Bronnikov-1973} by extending the works of Penney and others
\citep{Bronnikov-1972-Kiev,Zaitsev-1972,Penney-1969}. The success
of Bronnikov's work relies on a Weyl mapping that brings the action
into the Einstein frame, significantly simplifying the gravitation
sector while leaving the Maxwell field unaffected (a technical point
to be re-explained in this paper).\vskip4pt

The coupling of Brans-Dicke gravity with a massless Klein-Gordon field
has received relatively little attention in the existing literature
Unlike the Maxwell EM field, the Klein-Gordon field, upon the Weyl
mapping trick, becomes coupled with the `dilatonic' degree of freedom
of the BD action in the Einstein frame, thereby complicating the problem
\footnote{It should be noted that a closely related extension, allowing GR to
couple with a Maxwell field and a dilation field, has been investigated
by Cl\'ement et al \citep{Clement-2009}.}. At first, this technical complexity may seem like a steep price
to pay. Nevertheless, we shall demonstrate in this paper that this
obstacle can be overcome, leading to an exact analytical solution.\vskip4pt

Our paper is organized as follows. In Section \ref{sec:Recap} we
review the established ${\cal O}\left(1/\sqrt{\omega}\right)$ anomaly
for BD vacuum and Brans-Dicke-Maxwell electrovacuum. Section \ref{sec:BDKG-action}
introduces the Brans-Dicke-Klein-Gordon (BDKG) action, solves the
Einstein-frame field equations in the harmonic radial coordinate system,
and examines the solution at large $\omega$. Section \ref{sec:Formal-proof}
provides a formal proof of the observed ${\cal O}\left(1/\sqrt{\omega}\right)$
behavior for a generic matter source. Sections \ref{sec:gamma} and
\ref{sec:Physical-implications} derive the Eddington-Robertson-Schiff
parameters $\beta$ and $\gamma$, and discuss the physical implications
of our solution. Section \ref{sec:Conclusion} concludes and offers
an outlook on the implication of the ${\cal O}\left(1/\sqrt{\omega}\right)$
anomaly in the PPN formalism. Appendix \ref{sec:Field-equations}
derives the field equations of the BDKG action for the Einstein frame.
Appendix \ref{sec:Isotropic} expresses the solution in the isotropic
coordinate system.

\section{\label{sec:Recap}A recap of the ${\cal O}\left(1/\sqrt{\omega}\right)$
anomaly in Brans-Dicke vacuum}

Consider the Brans-Dicke (BD) action coupled with an external matter
source denoted by $\Psi$ \citep{BransDicke-1961,Brans-1962}:
\begin{align}
\mathcal{S}_{\text{BD}} & =\int d^{4}x\sqrt{-g}\Bigl[\frac{1}{16\pi}\Bigl(\phi\,\mathcal{R}-\frac{\omega}{\phi}g^{\mu\nu}\partial_{\mu}\phi\partial_{\nu}\phi\Bigr)\nonumber \\
 & \ \ \ \ \ \ \ \ \ \ \ \ \ \ \ \ \ \ \ \ \ \ \ \ \ +L^{(m)}\bigl(g_{\alpha\beta};\Psi\bigr)\Bigr]\label{eq:BD-action}
\end{align}
To respect Einstein's Equivalence Principle, the matter sector $L_{m}$
does not include the BD scalar field $\phi$ \citep{BransDicke-1961}.
The corresponding field equations are
\begin{align}
\mathcal{R}_{\mu\nu} & =\frac{\omega}{\phi^{2}}\nabla_{\mu}\phi\nabla_{\nu}\phi+\frac{1}{\phi}\nabla_{\mu}\nabla_{\nu}\phi+\frac{8\pi}{\phi}\Bigl[T_{\mu\nu}-\frac{\omega+1}{2\omega+3}g_{\mu\nu}T\Bigr]\label{eq:BD-eq-R}\\
\square\,\phi & =\frac{8\pi}{2\omega+3}T\label{eq:BD-eq-phi}
\end{align}
Here, the EMT of the matter sector is given by
\begin{equation}
T_{\mu\nu}:=\frac{-2}{\sqrt{-g}}\frac{\delta\,\bigl(\sqrt{-g}\,L^{(m)}(g_{\alpha\beta};\Psi)\bigr)}{\delta g^{\mu\nu}}\label{eq:EMT-def}
\end{equation}

A widely held belief, as popularized in Ref.$\,$\citep{Weinberg},
is that, since $\square\,\phi\simeq\mathcal{O}(1/\omega)$ by virtue
of Eq. \eqref{eq:BD-eq-phi}, the scalar field should exhibit the
following behavior
\begin{equation}
\phi=\bar{\phi}+\mathcal{O}\text{\ensuremath{\left(\frac{1}{\omega}\right)}}\label{eq:phi-behave}
\end{equation}
with $\bar{\phi}$ being a non-zero constant. The behavior in \eqref{eq:phi-behave}
is consistent with the (seemingly natural) following formal Taylor
expansion of the solution
\begin{align}
\phi & =\bar{\phi}\,\biggl(1+\frac{\stackrel{2}{\varphi}}{\omega}+\frac{\stackrel{4}{\varphi}}{\omega^{2}}+\dots\biggr)\label{eq:series-phi}\\
g_{\mu\nu} & =\ \stackrel{0}{g}_{\mu\nu}+\,\frac{\stackrel{2}{g}_{\mu\nu}}{\omega}+\,\frac{\stackrel{4}{g}_{\mu\nu}}{\omega^{2}}+\dots\label{eq:series-g}
\end{align}
where $\stackrel{0}{g}_{\mu\nu}$ is the \emph{limiting} (or ``remnant'')
metric as $\omega\rightarrow\infty$, and $\bigl\{\stackrel{2n}{\varphi},\,\stackrel{2n}{g}_{\mu\nu}\bigr\}$
are corrective components at order $1/\omega^{n}$, respectively.
As long as $\bar{\phi}>0$, the asymptotic behavior in \eqref{eq:series-phi}
results in
\begin{equation}
\frac{\omega}{\phi^{2}}\nabla_{\mu}\phi\nabla_{\nu}\phi\simeq\mathcal{O}\biggl(\frac{1}{\omega}\biggr)\ \text{and}\ \ \frac{1}{\phi}\nabla_{\mu}\nabla_{\nu}\phi\simeq\mathcal{O}\biggl(\frac{1}{\omega}\biggr)
\end{equation}
Consequently, with the aid of \eqref{eq:series-phi} and \eqref{eq:series-g},
Eq. \eqref{eq:BD-eq-R} to the leading order produces
\begin{equation}
\stackrel{0}{\mathcal{R}}_{\mu\nu}=\frac{8\pi}{\bar{\phi}}\left[\stackrel{0}{T}_{\mu\nu}-\frac{1}{2}\stackrel{0}{g}_{\mu\nu}\stackrel{0}{T}\right]\label{eq:a-1}
\end{equation}
in which $\stackrel{0}{\mathcal{R}}_{\mu\nu}$ is the Ricci tensor
calculated from $\stackrel{0}{g}_{\mu\nu}$ and $\stackrel{0}{T}_{\mu\nu}$
is the EMT evaluated at $\stackrel{0}{g}_{\mu\nu}$. By virtue of
Eq. \eqref{eq:a-1},\linebreak the ``remnant'' metric $\stackrel{0}{g}_{\mu\nu}$
thus satisfies the $\stackrel{0}{T}_{\mu\nu}-$sourced Einstein field
equation of General Relativity (GR), with the factor $8\pi/\bar{\phi}$
assuming the role of Newton's constant.\vskip8pt

The $\mathcal{O}(1/\omega)$ behavior of the BD scalar field has faced
challenges over the past thirty years. One concrete example central
to this debate is the Brans Class I exact solution---a static spherically
symmetric vacuum for BD gravity \citep{Brans-1962}. Previous investigations
of this solution \citep{Faraoni-1998,Faraoni-1999,Faraoni-2019,Romero-1993-a,Romero-1993-b,Romero-1993-c,Romero-1998,Romero-2021,Chauvineau-2003,Chauvineau-2007,Bhadra-2005,Bhadra-2002,Bhadra-2001,Banerjee-1997,Fabris-2019,Quiros-1999,Faraoni-2018}
have revealed that, in the limit of $\omega\rightarrow\infty$, (i)
its metric configuration does not converge to a vacuum of General
Relativity (GR), and (ii) its BD scalar field exhibits behavior in
the form
\begin{equation}
\phi=\bar{\phi}+\mathcal{O}\biggl(\frac{1}{\sqrt{\omega}}\biggr)\label{eq:a-2}
\end{equation}
in contrast to the conventionally believed form \eqref{eq:phi-behave}.\vskip4pt

The appearance of a square root in Expression \eqref{eq:a-2} may
seem perplexing at first, as the BD action \eqref{eq:BD-action} and
resulting field equations \eqref{eq:BD-eq-R}--\eqref{eq:BD-eq-phi}
do not inherently carry any non-analytical term. However, \emph{a
$\sqrt{\omega}$ term has been lurking underneath}, and it can be
exposed when expressing the action in the Einstein frame, as pointed
out by Faraoni and C\^ot\'e \citep{Faraoni-2019}. To see this,
through the following Weyl mapping
\begin{align}
\tilde{g}_{\mu\nu} & :=\phi\,g_{\mu\nu};\ \ \ \ \ \tilde{\phi}:=\sqrt{2\omega+3}\,\ln\frac{\phi}{\bar{\phi}}\label{eq:Weyl-mapping}
\end{align}
the BD action \eqref{eq:BD-action} transforms into a GR action minimally
coupled with a free scalar field $\tilde{\phi}$, viz.
\begin{equation}
\mathcal{S}_{\text{BD}}=\frac{1}{16\pi}\int d^{4}x\sqrt{-\tilde{g}}\Bigl[\tilde{\mathcal{R}}-\frac{1}{2}\tilde{g}^{\mu\nu}\partial_{\mu}\tilde{\phi}\partial_{\nu}\tilde{\phi}\Bigr]\label{eq:BD-EF-action}
\end{equation}
While this form of the BD action expressed in the Einstein frame is
``regular'' (i.e., not involving any non-analytical term), upon
returning to the Jordan frame, the $\sqrt{2\omega+3}$ participates,
per 
\begin{equation}
\phi=\bar{\phi}\,\exp\frac{\tilde{\phi}}{\sqrt{2\omega+3}}\simeq\bar{\phi}+\mathcal{O}\biggl(\frac{1}{\sqrt{\omega}}\biggr)
\end{equation}
hence resulting in the $\mathcal{O}(1/\sqrt{\omega})$ for the BD
scalar field. Consequently, this also leads to 
\begin{equation}
g_{\mu\nu}=\phi^{-1}\tilde{g}_{\mu\nu}=\bar{\phi}^{-1}\stackrel{0}{\tilde{g}_{\mu\nu}}+\mathcal{O}\biggl(\frac{1}{\sqrt{\omega}}\biggr)
\end{equation}
running counter to the series expansion \eqref{eq:series-g}. Moreover,
the ``remnant'' metric can be deduced to be proportional to $\tilde{g}_{\mu\nu}$,
viz.
\begin{equation}
\stackrel{0}{g}_{\mu\nu}=\bar{\phi}^{-1}\,\stackrel{0}{\tilde{g}_{\mu\nu}}\label{eq:link-g}
\end{equation}
Moreover, it is important to note that, by virtue of the action \eqref{eq:BD-EF-action},
$\tilde{g}_{\mu\nu}$ is \emph{not} a vacuum of GR. Rather, it is
a configuration of GR sourced by a free massless scalar field (i.e.,
$\tilde{\phi}$). For the static spherically symmetric setup, $\tilde{g}_{\mu\nu}$
is known to be the FJNW spacetime \citep{Bergmann-1957,Wyman-1981,Fisher-1948,Janis-1968,Buchdahl-BD-1972}
which deviates from the Schwarzschild spacetime.\vskip4pt

This exposition underscores an essential point, however: \emph{the
square root term $\sqrt{2\omega+3}$ is intrinsic to the gravitational
sector itself.} That is to say, \emph{the $\sqrt{2\omega+3}$ term
has nothing to do with the matter sector}. Therefore, it is natural
to anticipate that the $\sqrt{2\omega+3}$ term should leave its footprints
in a Brans-Dicke theory in general, \emph{regardless} of the trace
of the EMT of the source. This anticipation constitutes the central
theme of our present paper, wherein we aim to rigorously substantiate
and validate this expectation.\vskip4pt

It should be noted that prior challenges to the $\mathcal{O}(1/\omega)$
behavior in BD gravity have predominantly focused on scenarios involving
vacuum or electrovacuum \citep{Romero-1993-a,Romero-1993-b,Romero-1993-c,Romero-1998,Romero-2021,Faraoni-1998,Faraoni-1999,Faraoni-2019,Bhadra-2001,Bhadra-2002,Bhadra-2005,Fabris-2019}.
Authors have been cautious about extending these challenges to situations
where the EMT of the matter source possesses a \emph{non-vanishing}
trace. It has also been contended that while the convergence to GR
and the $\mathcal{O}(1/\omega)$ behavior may falter for traceless
matter, they remain valid for trace-carrying matter \citep{Faraoni-1998,Faraoni-1999}.
An exception was one of the current authors who suggested the generality
of the $O(1/\sqrt{\omega})$ behavior for non-vanishing EMT trace
\citep{Chauvineau-2003,Chauvineau-2007}.

\section{\label{sec:BDKG-action}Brans-Dicke-Klein-Gordon action: New exact
solution and $\mathcal{O}\left(1/\sqrt{\omega}\right)$ anomaly}

The central focus of our paper is the Brans-Dicke-Klein-Gordon (BDKG)
action, where the Brans-Dicke sector is minimally coupled with a free
massless Klein-Gordon (KG) scalar field $\Psi$, given as \footnote{For the sake of clarity, we shall drop the $1/(16\pi)$ factor that
is conventionally associated with the gravitation sector.}
\begin{equation}
\sqrt{-g}\biggl[\phi\,\mathcal{R}-\frac{\omega}{\phi}g^{\mu\nu}\partial_{\mu}\phi\partial_{\nu}\phi-g^{\mu\nu}\partial_{\mu}\Psi\partial_{\nu}\Psi\biggr]\label{eq:BDKG-action}
\end{equation}
The EMT of the KG field is $T_{\mu\nu}:=\partial_{\mu}\Psi\partial_{\nu}\Psi-\frac{1}{2}g_{\mu\nu}(\nabla\Psi)^{2}$
which has a non-vanishing trace $T=-(\nabla\Psi)^{2}$. The field
equations are
\begin{align}
\mathcal{R}_{\mu\nu} & =\frac{\omega}{\phi^{2}}\nabla_{\mu}\phi\nabla_{\nu}\phi+\frac{1}{\phi}\nabla_{\mu}\nabla_{\nu}\phi\nonumber \\
 & +\frac{1}{\phi}\left[\partial_{\mu}\Psi\partial_{\nu}\Psi-g_{\mu\nu}\frac{(\nabla\Psi)^{2}}{2(2\omega+3)}\right]\label{eq:eqn-Ricci}\\
\square\,\phi & =-\frac{(\nabla\Psi)^{2}}{2\omega+3}\label{eq:eqn-phi}\\
\square\,\Psi & =0\label{eq:eqn-Psi}
\end{align}
These equations are invariant through the two transformations: (1)
$\Psi\rightarrow-\Psi$ and (2) $\Psi\rightarrow\Psi+\text{constant}$.

\subsection{Difficulties in the Jordan frame}

This subsection reveals the technical difficulties associated with
the Jordan frame, and the need to move to the Einstein frame. To see
this, let us consider a static spherically symmetric metric in the
harmonic radial coordinate, viz.
\begin{equation}
ds^{2}=-e^{2\gamma(u)}dt^{2}+e^{-2\gamma(u)}\left[e^{2\alpha(u)}du^{2}+e^{\alpha(u)}d\Omega^{2}\right]
\end{equation}
which yields
\begin{align}
\sqrt{-g} & =e^{-2\gamma(u)+2\alpha(u)}\sin\theta\\
\sqrt{-g}\,g^{uu} & =\sin\theta
\end{align}
In the harmonic radial coordinate, Eqs. \eqref{eq:eqn-phi} and \eqref{eq:eqn-Psi}
are straightforward to handle. First, Eq. \eqref{eq:eqn-Psi} gives
(without loss of generality) \footnote{Recall that for a scalar field $\phi$, $\square\phi=(\sqrt{-g})^{-1}\partial_{\mu}(\sqrt{-g}g^{\mu\nu}\partial_{\nu}\phi)$. }
\begin{equation}
\Psi(u)=\Psi_{*}u\label{eq:tmp-0}
\end{equation}
with $\Psi_{*}$ being an integration constant. Next, from \eqref{eq:eqn-phi}
and \eqref{eq:tmp-0}, we have
\begin{equation}
\phi(u)=-\frac{\Psi_{*}^{2}}{2(2\omega+3)}u^{2}+Nu+v_{0}
\end{equation}
with $N$ and $v_{0}$ being integration constants. This simplicity
is misleading, however. There is ``a \emph{steep} price to pay'':
the operator $\nabla_{\mu}\nabla_{\nu}\phi$ in Eq. \eqref{eq:eqn-Ricci}
typically involves cumbersome Christoffel symbols. After full calculation
of the Christoffel symbols and the Ricci tensor, the field equations
leads to
\begin{equation}
\left\{ \begin{array}{l}
\gamma''=-\frac{1}{\phi}\gamma'\phi'+\frac{1}{\phi}\frac{\Psi'^{2}}{2(2\omega+3)}\\
-\alpha''+\frac{1}{2}\alpha'^{2}+\gamma''-2\gamma'^{2}\\
\ \ \ \ =\frac{\omega}{\phi^{2}}\phi'^{2}+\frac{1}{\phi}\left(\phi''-(\alpha'-\gamma')\phi'\right)+\frac{1}{\phi}\frac{4\omega+5}{2(2\omega+3)}\Psi'^{2}\\
\ e^{\alpha}+\gamma''-\frac{\alpha''}{2}=-\frac{1}{\phi}\left(\gamma'-\frac{\alpha'}{2}\right)\phi'+\frac{1}{\phi}\frac{\Psi'^{2}}{2(2\omega+3)}
\end{array}\right.
\end{equation}
Solving for $\alpha(u)$ and $\gamma(u)$ from these equations proves
to be unwieldy, rendering the Jordan frame unsuitable for obtaining
a metric in analytical form. Following the approach introduced by
Bronnikov, we will first transition into the Einstein frame, a strategy
that has demonstrated advantages \citep{Bronnikov-1973}.

\subsection{Mapping from Jordan frame to Einstein frame}

For the sake of illustration, instead of \eqref{eq:BDKG-action},
we shall consider a BD action minimally coupled with a massless KG
field $\Psi$
\begin{equation}
\sqrt{-g}\Bigl[\phi\,\mathcal{R}-\frac{\omega}{\phi}g^{\mu\nu}\partial_{\mu}\phi\partial_{\nu}\phi-g^{\mu\nu}\partial_{\mu}\Psi\partial_{\nu}\Psi\Bigr]\,.\label{eq:full-action}
\end{equation}
Let us define a new field $\theta$ per
\begin{equation}
\phi:=e^{\theta}
\end{equation}
and make a Weyl rescaling
\begin{align}
\tilde{g}_{\mu\nu} & =e^{\theta}g_{\mu\nu},\ \ \ \tilde{g}^{\mu\nu}=e^{-\theta}g^{\mu\nu},\ \ \ \sqrt{-\tilde{g}}=e^{2\theta}\sqrt{-g}
\end{align}
the Ricci scalar transforms as \citep{useful}
\begin{equation}
\mathcal{R}=e^{\theta}\Bigl[\tilde{\mathcal{R}}+3\tilde{\square}\,\theta-\frac{3}{2}(\tilde{\nabla}\theta)^{2}\Bigr]
\end{equation}
These expressions give (modulo a total derivative term in the first
identity below)
\begin{align}
\sqrt{-g}\phi\mathcal{R} & =\sqrt{-\tilde{g}}\left[\tilde{\mathcal{R}}-\frac{3}{2}\tilde{g}^{\mu\nu}\partial_{\mu}\theta\partial_{\nu}\theta\right]\\
\sqrt{-g}\frac{\omega}{\phi}g^{\mu\nu}\partial_{\mu}\phi\partial_{\nu}\phi & =\sqrt{-\tilde{g}}\omega\tilde{g}^{\mu\nu}\partial_{\mu}\theta\partial_{\nu}\theta\\
\sqrt{-g}g^{\mu\nu}\partial_{\mu}\Psi\partial_{\nu}\Psi & =\sqrt{-\tilde{g}}\tilde{g}^{\mu\nu}e^{-\theta}\partial_{\mu}\Psi\partial_{\nu}\Psi\label{eq:transforming-Psi}
\end{align}
The full action \eqref{eq:full-action} becomes
\begin{equation}
\sqrt{-\tilde{g}}\Bigl[\tilde{\mathcal{R}}-\left(\omega+\frac{3}{2}\right)\tilde{g}^{\mu\nu}\partial_{\mu}\theta\partial_{\nu}\theta-e^{-\theta}\tilde{g}^{\mu\nu}\partial_{\mu}\Psi\partial_{\nu}\Psi\Bigr]\label{eq:EKGD-action}
\end{equation}
Crucially, the KG field now forms a coupling with the BD scalar field,
where the latter takes on the role of a ``dilaton'', represented by
the term $e^{-\theta}$. This introduces a level of complexity for
the KG field.\vskip8pt

The field equations for the action \eqref{eq:EKGD-action} in the
\emph{Einstein frame} are derived to be (see Appendix \ref{sec:Field-equations})
\begin{equation}
\tilde{\mathcal{R}}_{\mu\nu}=\left(\omega+3/2\right)\partial_{\mu}\theta\partial_{\nu}\theta+e^{-\theta}\partial_{\mu}\Psi\partial_{\nu}\Psi\label{eq:e-3}
\end{equation}
\begin{align}
(2\omega+3)\,e^{-\theta}\,\tilde{\square}\,\theta & =-(e^{-\theta}\tilde{\nabla}^{\mu}\Psi)\,(e^{-\theta}\tilde{\nabla}_{\mu}\Psi)\label{eq:e-4}\\
\tilde{\nabla}^{\mu}\left(e^{-\theta}\tilde{\nabla}_{\mu}\Psi\right) & =0\label{eq:e-5}
\end{align}

\subsection{\label{subsec:Solving-scalar}Solving for the scalar fields}

Assuming stationarity and spherical symmetry, we shall work in the
harmonic radial coordinate \footnote{A harmonic radial coordinates satisfies the identity $\sqrt{-g}g^{11}=1$.}
in the Einstein frame, viz.
\begin{align}
d\tilde{s}^{2} & =-e^{2\gamma(u)}dt^{2}+e^{-2\gamma(u)}\left[e^{2\alpha(u)}du^{2}+e^{\alpha(u)}d\Omega^{2}\right]\label{eq:gauge}\\
\sqrt{-\tilde{g}} & =e^{-2\gamma(u)+2\alpha(u)}\sin\theta\\
\sqrt{-\tilde{g}}\tilde{g}^{uu} & =\sin\theta
\end{align}
The following identities will be useful
\begin{align}
\tilde{g}^{\mu\nu}\tilde{\Gamma}_{\mu\nu}^{\lambda} & =\frac{-1}{\sqrt{-\tilde{g}}}\partial_{\mu}(\sqrt{-\tilde{g}}\,\tilde{g}^{\lambda\mu})\label{eq:ident-1}\\
\tilde{\square}\,\theta & =\frac{1}{\sqrt{-\tilde{g}}}\partial_{\mu}(\sqrt{-\tilde{g}}\tilde{g}^{\mu\nu}\partial_{\nu}\Psi)\,.
\end{align}
The left hand side of Eq. \eqref{eq:e-5} becomes
\begin{align}
 & \tilde{\nabla}^{\mu}(e^{-\theta}\tilde{\nabla}_{\mu}\Psi)\nonumber \\
 & =\tilde{g}^{\mu\nu}\partial_{\nu}(e^{-\theta}\partial_{\mu}\Psi)+\frac{1}{\sqrt{-\tilde{g}}}\partial_{\mu}(\sqrt{-\tilde{g}}\,\tilde{g}^{\lambda\mu})e^{-\theta}\partial_{\lambda}\Psi\\
 & =e^{2\gamma-2\alpha}\left(e^{-\theta}\Psi'\right)'
\end{align}
in which the identity \eqref{eq:ident-1} has been employed. Equation
\eqref{eq:e-5} then yields
\begin{equation}
\Psi'=\Psi_{*}e^{\theta}=\Psi_{*}\phi\label{eq:b-0}
\end{equation}
with $\Psi_{*}$ being an integration constant. With the aid of \eqref{eq:b-0},
Eq. \eqref{eq:e-4} yields
\begin{align}
\theta'' & =-\frac{\Psi_{*}^{2}}{2\omega+3}e^{\theta}\label{eq:b-1}
\end{align}
Multiplying both sides with $d\theta=\theta'du$ and integrating
\begin{equation}
(\theta')^{2}+\frac{2\Psi_{*}^{2}}{2\omega+3}e^{\theta}=4\kappa\in\mathbb{R}\label{eq:b-2}
\end{equation}
which has a solution
\begin{equation}
\theta(u)=\begin{cases}
\ {\displaystyle \ln\frac{2(2\omega+3)\,\kappa}{\Psi_{*}^{2}\cosh^{2}\sqrt{\kappa}(u+u_{0})}} & (\omega>-3/2,\,\kappa\geqslant0)\\
\ {\displaystyle \ln\frac{-2(2\omega+3)\,\kappa}{\Psi_{*}^{2}\sinh^{2}\sqrt{\kappa}(u+u_{0})}} & (\omega<-3/2,\,\kappa\in\mathbb{R})
\end{cases}\label{eq:b-3}
\end{equation}
or
\begin{equation}
\phi(u)=\begin{cases}
\ {\displaystyle \frac{2(2\omega+3)\,\kappa}{\Psi_{*}^{2}\,\cosh^{2}\sqrt{\kappa}(u+u_{0})}} & (\omega>-3/2,\,\kappa\geqslant0)\\
\ {\displaystyle \frac{-2(2\omega+3)\,\kappa}{\Psi_{*}^{2}\,\sinh^{2}\sqrt{\kappa}(u+u_{0})}} & (\omega<-3/2,\,\kappa\in\mathbb{R})
\end{cases}\label{eq:b-4}
\end{equation}
from which, $\phi(u)\propto e^{-2\sqrt{\kappa}\left|u\right|}\rightarrow0$
as $u\rightarrow\pm\infty$.\vskip8pt

\paragraph{\bf{\emph{Recasting the scalar fields using the limit of $\Psi_{*}\rightarrow0$:}}}

\begin{figure*}[!t]
\noindent \begin{centering}
\includegraphics{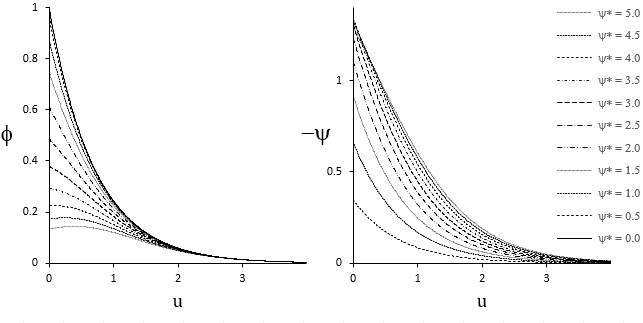}
\par\end{centering}
\caption{\label{fig:limit}$\phi(u)$ and $-\Psi(u)$ as functions of $u$,
for various values of $\Psi_{*}$ as indicated in legend. In these
plots, $\omega=0.3$, $\kappa\equiv0.5$, $\phi_{0}\equiv1$.}
\end{figure*}

For $\Psi_{*}=0$, Eq. \eqref{eq:b-2} yields two affine solutions,
$\theta(u)=\pm2\sqrt{\kappa}(u-u_{0})$ or
\begin{equation}
\phi(u)=\phi_{0}\,e^{\pm2\sqrt{\kappa}u}\label{eq:c-1}
\end{equation}
We may not directly send $\Psi_{*}$ to zero in Eq. \eqref{eq:b-4},
however. Some delicate handling is needed. At each value of $\Psi_{*}$,
we have the liberty to choose $u_{0}$ in Eq. \eqref{eq:b-4} such
that the asymptotic behavior of $\phi(u)$ for large $u$ matches
Eq. \eqref{eq:c-1}. In particular, for $\omega>-3/2$, (note: $\kappa\geqslant0$)
consider
\begin{equation}
{\displaystyle \frac{2(2\omega+3)\,\kappa}{\Psi_{*}^{2}\,\cosh^{2}\sqrt{\kappa}(u+u_{0})}={\displaystyle \frac{8(2\omega+3)\,\kappa}{\Psi_{*}^{2}e^{2\sqrt{\kappa}u_{0}}}\frac{e^{-2\sqrt{\kappa}u}}{\left(1+e^{-2\sqrt{\kappa}(u+u_{0})}\right)^{2}}}}\label{eq:c-2}
\end{equation}
We then set
\begin{equation}
\frac{8(2\omega+3)\kappa}{\Psi_{*}^{2}e^{2\sqrt{\kappa}u_{0}}}\equiv\phi_{0}\in\mathbb{R}^{+}
\end{equation}
and bring Eq. \eqref{eq:c-2} to
\begin{equation}
\frac{2(2\omega+3)\,\kappa}{\Psi_{*}^{2}\,\cosh^{2}\sqrt{\kappa}(u+u_{0})}=\frac{\phi_{0}\,e^{-2\sqrt{\kappa}u}}{\left(1+\frac{\Psi_{*}^{2}\phi_{0}}{8(2\omega+3)\kappa}e^{-2\sqrt{\kappa}u}\right)^{2}}
\end{equation}
Similarly for $\omega<-3/2$ (with $e^{-2\sqrt{\kappa}u_{0}}=\frac{-\Psi_{*}^{2}\phi_{0}}{8(2\omega+3)\kappa}$
if $\kappa\geqslant0$) \footnote{Note: the case of $\omega<-3/2$ and $\kappa<0$ does not have a limit
to the pure BD.}
\begin{align}
\frac{-2(2\omega+3)\,\kappa}{\Psi_{*}^{2}\,\sinh^{2}\sqrt{\kappa}(u+u_{0})} & =\frac{\phi_{0}\,e^{-2\sqrt{\kappa}u}}{\left(1+\frac{\Psi_{*}^{2}\phi_{0}}{8(2\omega+3)\kappa}e^{-2\sqrt{\kappa}u}\right)^{2}}
\end{align}

In summary, we have chosen a different parametrization for $\phi(u)$.
We restrict ourselves to the case of $\kappa\geqslant0$ (but no restriction
on $\omega$, except that $\omega\neq-3/2$). We have set
\begin{equation}
u_{0}=-\frac{1}{2\sqrt{\kappa}}\ln\frac{\Psi_{*}^{2}\phi_{0}}{8\left|2\omega+3\right|\kappa}\label{eq:def-u0}
\end{equation}
and brought Eq. \eqref{eq:b-4} to the following form
\begin{equation}
\phi(u)=\frac{\phi_{0}\,e^{-2\sqrt{\kappa}u}}{\left(1+\frac{\Psi_{*}^{2}\phi_{0}}{8(2\omega+3)\kappa}e^{-2\sqrt{\kappa}u}\right)^{2}}\label{eq:phi}
\end{equation}
which gracefully recovers the pure BD case when $\Psi_{*}\rightarrow0$,
viz. $\phi(u)=\phi_{0}e^{-2\sqrt{\kappa}u}$.\vskip4pt

By virtue of $\Psi(u)=\Psi_{*}\int du\,\phi(u)$ (choosing $\Psi(u=+\infty)=0$):
\begin{align}
\Psi(u) & =\frac{4(2\omega+3)\sqrt{\kappa}}{\Psi_{*}\left(1+\frac{\Psi_{*}^{2}\phi_{0}}{8(2\omega+3)\kappa}e^{-2\sqrt{\kappa}u}\right)}-\frac{4(2\omega+3)\sqrt{\kappa}}{\Psi_{*}}\\
 & =-\frac{\Psi_{*}\phi_{0}}{2\sqrt{\kappa}}\frac{e^{-2\sqrt{\kappa}u}}{1+\frac{\Psi_{*}^{2}\phi_{0}}{8(2\omega+3)\kappa}e^{-2\sqrt{\kappa}u}}\label{eq:Psi}
\end{align}
In the limit of $\Psi_{*}\rightarrow0$, the KG field disappears as
expected. Figure Fig. \ref{fig:limit} illustrates the functions $\phi(u)$
and $\Psi(u)$ for various values of $\Psi_{*}$.

\subsection{\label{subsec:Solving-metric}Solving for the metric components}

A great virtue of the Einstein frame over the Jordan frame is that
the field equation \eqref{eq:e-3} for the metric tensor does not
involve the term $\nabla_{\mu}\nabla_{\nu}\phi$ which typically involves
the cumbersome Christoffel symbols. Moreover, the $tt-$ and $\theta\theta-$components
of Eq. \eqref{eq:EF-eqn-R} are straightforward to solve as they are
decoupled from the scalar fields.\vskip4pt

Recall that for the metric expressed in the harmonic radial coordinate
\eqref{eq:gauge}, the Ricci tensor components are
\begin{align}
\tilde{\mathcal{R}}_{tt} & =e^{4\gamma-2\alpha}\gamma''\label{eq:Ricci-00}\\
\tilde{\mathcal{R}}_{uu} & =-\alpha''+\frac{1}{2}\alpha'^{2}+\gamma''-2\gamma'^{2}\label{eq:Ricci-11}\\
\tilde{\mathcal{R}}_{\theta\theta} & =1+e^{-\alpha}\Bigl(\gamma''-\frac{1}{2}\alpha''\Bigr)\label{eq:Ricci-22}
\end{align}
Invoking the $tt-$component of Eq. \eqref{eq:e-3} and using \eqref{eq:Ricci-00}
leads to
\begin{equation}
\gamma(u)=-a\,(u+u_{1})\,.\label{eq:gamma-solved}
\end{equation}
Invoking the $\theta\theta-$component of Eq. \eqref{eq:e-3} and
using \eqref{eq:Ricci-22} and \eqref{eq:gamma-solved} yields (without
loss of generality)
\begin{align}
\alpha''=2e^{\alpha}\ \ \Longrightarrow\ \ \alpha(u) & =\frac{\chi}{\sinh^{2}\sqrt{\chi}\,u}\,.\label{eq:alpha-solved}
\end{align}
We should note that for $\chi\in\mathbb{R^{-}}$ the $sinh$ function
is expressible as a $sin$ function per
\begin{align}
\frac{\sinh\sqrt{\chi}\,u}{\sqrt{\chi}} & =\frac{\sin\sqrt{-\chi}\,u}{\sqrt{-\chi}}\,,
\end{align}
and for $\chi=0$, the $sinh$ function admits the following limit
\begin{equation}
\lim_{\chi\rightarrow0}\frac{\sinh\sqrt{\chi}\,u}{\sqrt{\chi}}=u\,.
\end{equation}

Next, with the aid of \eqref{eq:gamma-solved} and \eqref{eq:alpha-solved},
Eq. \eqref{eq:Ricci-11} produces
\begin{equation}
\tilde{\mathcal{R}}_{uu}=2\left(\chi-a^{2}\right)\label{eq:f-1}
\end{equation}
On the other hand, the right-hand-side of the $uu-$component of Eq.
\eqref{eq:EF-eqn-R} is
\begin{align}
\left(\omega+3/2\right)\partial_{\mu}\theta\partial_{\nu}\theta+e^{-\theta}\partial_{\mu}\Psi\partial_{\nu}\Psi & =2(2\omega+3)\,\kappa\label{eq:f-2}
\end{align}

From Eqs. \eqref{eq:f-1} and \eqref{eq:f-2}, the $uu-$component
of Eq. \eqref{eq:EF-eqn-R} becomes a ``constraint'' among the parameters,
given by
\begin{equation}
\chi=a^{2}+(2\omega+3)\,\kappa\,.\label{eq:constraint}
\end{equation}
Therefore, the metric in the \emph{Einstein frame} is
\begin{align}
d\tilde{s}^{2} & =-e^{-2a(u+u_{1})}dt^{2}\nonumber \\
 & +\frac{e^{2a(u+u_{1})}\chi^{2}}{\sinh^{4}\sqrt{\chi}\,u}du^{2}+\frac{e^{2a(u+u_{1})}\chi}{\sinh^{2}\sqrt{\chi}\,u}d\Omega^{2}\,.\label{eq:EF-metric}
\end{align}
\vskip4pt

\paragraph{\bf{\emph{Removal of $u_{1}$:}}}

The parameter $u_{1}$ in \eqref{eq:EF-metric} can be removed via
rescaling of variables and parameters, per
\begin{align}
u & \rightarrow e^{au_{1}}u;\ \ \ t\rightarrow e^{au_{1}}t\\
\chi & \rightarrow e^{-2au_{1}}\chi;\ \ \ a\rightarrow e^{-au_{1}}a\\
\kappa & \rightarrow e^{-2au_{1}}\kappa;\ \ \ \Psi_{*}\rightarrow e^{-au_{1}}\Psi_{*}
\end{align}
Let us remark that Eqs. \eqref{eq:phi}, \eqref{eq:Psi}, and the
``constraint'' \eqref{eq:constraint} are unaffected.

\subsection{\label{subsec:BDKG-solution}Exact solution of BDKG action in Jordan
frame}

Moving back to the Jordan frame, the final solution is (with the convention
of $a>0$) \footnote{We reemphasize that when $\chi<0$ the function $\frac{\sqrt{\chi}}{\sinh\sqrt{\chi}u}$
is \emph{numerically} equal to $\frac{\sqrt{-\chi}}{\sin\sqrt{-\chi}u}$.}
\begin{equation}
ds^{2}=\frac{1}{\phi(u)}\Bigl[-e^{-2au}dt^{2}+e^{2au}\left(\frac{\chi^{2}\,du^{2}}{\sinh^{4}\sqrt{\chi}\,u}+\frac{\chi\,d\Omega^{2}}{\sinh^{2}\sqrt{\chi}\,u}\right)\Bigr]\,,\label{eq:BDKG-solution}
\end{equation}
the BD scalar field is
\begin{equation}
\phi(u)={\displaystyle \frac{\phi_{0}\,\exp\left(-2\sqrt{\frac{\chi-a^{2}}{2\omega+3}}u\right)}{\left(1+\frac{\Psi_{*}^{2}\phi_{0}}{8(\chi-a^{2})}\exp\left(-2\sqrt{\frac{\chi-a^{2}}{2\omega+3}}u\right)\right)^{2}}}\,,\label{eq:BD-scalar-solution}
\end{equation}
and the KG field is
\begin{equation}
\Psi(u)=-\frac{\Psi_{*}\phi_{0}}{2\sqrt{\frac{\chi-a^{2}}{2\omega+3}}}\frac{\exp\left(-2\sqrt{\frac{\chi-a^{2}}{2\omega+3}}u\right)}{1+\frac{\Psi_{*}^{2}\phi_{0}}{8(\chi-a^{2})}\exp\left(-2\sqrt{\frac{\chi-a^{2}}{2\omega+3}}u\right)}\,.\label{eq:KG-scalar-solution}
\end{equation}
The two scalar fields expressed above are obviously related, in accordance
with Eq. \eqref{eq:b-0}. The integration constants entering this
solution are $\{a,\,\chi,\,\phi_{0},\,\Psi_{*}\}$.\vskip4pt

Finally, we have verified by \emph{direct inspection} that the solution
given in \eqref{eq:BDKG-solution}--\eqref{eq:KG-scalar-solution}
satisfies the field equations \eqref{eq:eqn-Ricci}--\eqref{eq:eqn-Psi}
of the original BDKG action \eqref{eq:BDKG-action} in the \emph{Jordan
frame.}

\subsection*{In ``reciprocal'' coordinate}

The final solution \eqref{eq:BDKG-solution}--\eqref{eq:KG-scalar-solution}
can also be expressed in a more familiar coordinate. Define 
\begin{equation}
r:=1/u\,.\label{eq:reciprocal-coord}
\end{equation}
The limit $u\rightarrow0$ would correspond to the spatial infinity
$r\rightarrow\infty$. The metric \eqref{eq:BDKG-solution} becomes
\begin{align}
ds^{2} & =\frac{1}{\phi(r)}\biggl[-e^{-\frac{2a}{r}}dt^{2}+e^{\frac{2a}{r}}\left(\frac{\sqrt{\chi}/r}{\sinh\sqrt{\chi}/r}\right)^{4}dr^{2}\nonumber \\
 & \ \ \ \ \ \ \ \ \ \ \ \ \ \ \ \ \ +e^{\frac{2a}{r}}\left(\frac{\sqrt{\chi}/r}{\sinh\sqrt{\chi}/r}\right)^{2}r^{2}d\Omega^{2}\biggr]\,,\label{eq:reciprocal-metric}
\end{align}
the BD scalar field \eqref{eq:BD-scalar-solution} is
\begin{equation}
\phi(r)={\displaystyle \frac{\phi_{0}\,e^{-\frac{2\sqrt{\kappa}}{r}}}{\left(1+\frac{\Psi_{*}^{2}\phi_{0}}{8(\chi-a^{2})}e^{-\frac{2\sqrt{\kappa}}{r}}\right)^{2}}}\,,\label{eq:reciprocal-BD}
\end{equation}
and the KG field \eqref{eq:KG-scalar-solution} is
\begin{equation}
\Psi(r)=-\frac{\Psi_{*}\phi_{0}}{2\sqrt{\kappa}}\frac{e^{-\frac{2\sqrt{\kappa}}{r}}}{1+\frac{\Psi_{*}^{2}\phi_{0}}{8(\chi-a^{2})}e^{-\frac{2\sqrt{\kappa}}{r}}}\,,\label{eq:reciprocal-KG}
\end{equation}
in which $\kappa=(\chi-a^{2})/(2\omega+3)$. When $\Psi_{*}=0$, the
$\Psi$ field disappears, whereas Expressions \eqref{eq:reciprocal-metric}
and \eqref{eq:reciprocal-BD} reproduce the four classes of the Brans
solution (Class I if $\chi>0$, Class II if $\chi<0$, Classes III
and IV if $\chi=0$).

\subsection{\label{sec:Behaviors}The $\mathcal{O}\left(1/\sqrt{\omega}\right)$
anomaly in BDKG}

The exact solution \eqref{eq:reciprocal-metric}--\eqref{eq:reciprocal-KG}
stands handy for us to examine the large $\omega$ limit.\vskip4pt

First of all, the static spherically symmetric metric for BDKG theory
is conformally equivalent to the FJNW spacetime 
\begin{align}
ds_{\text{FJNW}}^{2} & =-e^{-\frac{2a}{r}}dt^{2}+e^{\frac{2a}{r}}\left(\frac{\sqrt{\chi}/r}{\sinh\sqrt{\chi}/r}\right)^{4}dr^{2}\nonumber \\
 & \ \ \ \ \ \ \ \ \ \ \ +e^{\frac{2a}{r}}\left(\frac{\sqrt{\chi}/r}{\sinh\sqrt{\chi}/r}\right)^{2}r^{2}d\Omega^{2}\label{eq:FJNW}
\end{align}
in which $\chi$ and $a$ are \emph{two independent parameters}. Only
when $\chi=a^{2}\in\mathbb{R}^{+}$ does the metric \eqref{eq:FJNW}
become the Schwarzschild metric.\vskip4pt

For a given pair of $\{\chi,\,a\}$, the metric of BDKG gravity and
the FJNW metric would share a common causal structure. For a given
causal structure, viz. keeping $\chi$ and $a$ fixed, the dependence
on $\omega$ is encapsulated in the conformal factor, viz. $1/\phi(r)$.\vskip8pt

\subsubsection{The $\omega\rightarrow\infty$ behavior of the scalar fields}

The Taylor expansion of Eq. \eqref{eq:reciprocal-BD} in terms of
$1/\sqrt{\omega}$ produces
\begin{align}
\phi(u) & =\frac{\phi_{0}}{\left(1+\frac{\Psi_{*}^{2}\phi_{0}}{8(\chi-a^{2})}\right)^{2}}\nonumber \\
 & -\frac{\phi_{0}\sqrt{2(\chi-a^{2})}\left(1-\frac{\Psi_{*}^{2}\phi_{0}}{8(\chi-a^{2})}\right)}{\left(1+\frac{\Psi_{*}^{2}\phi_{0}}{8(\chi-a^{2})}\right)^{3}}\frac{1}{\sqrt{\omega\,}r}+\mathcal{O}\left(\frac{1}{\omega}\right).\label{eq:taylor-phi}
\end{align}
That is to say, apart from the constant term $\bar{\phi}:=\phi_{0}/\left(1+\frac{\Psi_{*}^{2}\phi_{0}}{8(\chi-a^{2})}\right)^{2}$,
the sub-leading contribution to the $\phi$ field is in the order
of $1/\sqrt{\omega}$. Similarly, the Taylor expansion of Eq. \eqref{eq:reciprocal-KG}
yields
\begin{align}
\Psi(r) & =-\frac{\Psi_{*}\phi_{0}\,\sqrt{\omega}}{2\sqrt{2(\chi-a^{2})}\left(1+\frac{\Psi_{*}^{2}\phi_{0}}{8(\chi-a^{2})}\right)}\nonumber \\
 & +\frac{\Psi_{*}\phi_{0}}{\left(1+\frac{\Psi_{*}^{2}\phi_{0}}{8(\chi-a^{2})}\right)^{2}}\frac{1}{r}+\mathcal{O}\left(\frac{1}{\sqrt{\omega}}\right)\label{eq:taylor-Psi}
\end{align}
Note that the leading term is a constant; it induces no physical effect
and can be dropped. \vskip8pt

\subsubsection{The $\omega\rightarrow\infty$ ``remnant'' metric}

With $\chi$ and $a$ held fixed, $\kappa$ behaves as $1/\sqrt{|\omega|}$.
Upon sending $\omega$ to infinity, the BD scalar field thus becomes
a constant field. However, the limiting metric is not a Schwarzschild metric, but an FJNW one. In the harmonic radial coordinate
\begin{equation}
ds^{2}=\frac{1}{\bar{\phi}}\Bigl[-e^{-2au}dt^{2}+e^{2au}\left(\frac{\chi^{2}\,du^{2}}{\sinh^{4}\sqrt{\chi}\,u}+\frac{\chi\,d\Omega^{2}}{\sinh^{2}\sqrt{\chi}\,u}\right)\Bigr]\,,\label{eq:BDKG-limit}
\end{equation}
whereas $\phi(u)\equiv\bar{\phi}:=\phi_{0}/\left(1+\frac{\Psi_{*}^{2}\phi_{0}}{8(\chi-a^{2})}\right)^{2}$
and $\Psi(u)\equiv\text{const}$. Remark: a constant $\Psi$ has no
physical impact at all (while a constant $\phi$ is physically present
in the field equations). Expression \eqref{eq:BDKG-limit} is the
metric part of the FJNW solution of GR sourced by a scalar field
\begin{equation}
S_{\text{BD}}=\int d^{4}x\sqrt{-g}\,\phi_{0}\biggl(\mathcal{R}-g^{\mu\nu}\nabla_{\mu}\varphi\nabla_{\nu}\varphi\biggr)\,.
\end{equation}

\section{\label{sec:Formal-proof}Formal proof of the ${\cal O}\left(1/\sqrt{\omega}\right)$
anomaly for Brans--Dicke action sourced by any matter field}

In this section, we consider a BD action coupled with a matter field
$\Psi$ in the form
\begin{align}
\mathcal{S}_{\text{generic}} & =\int d^{4}x\sqrt{-g}\Bigl[\phi\,\mathcal{R}-\frac{\omega}{\phi}g^{\mu\nu}\partial_{\mu}\phi\partial_{\nu}\phi\nonumber \\
 & \ \ \ \ \ \ \ \ \ \ \ \ \ \ \ \ \ \ \ \ \ \ \ \ +L^{(m)}\bigl(\{g_{\alpha\beta},\phi\};\Psi\bigr)\Bigr]\label{eq:action-generic}
\end{align}
The matter sector can be of \emph{generic nature}; the $\Psi$ field
can be scalar, spinor, massive, and so forth. It may also couple with
the BD scalar field and/or couple with the metric tensor in a non-minimal
fashion. The EMT of the matter sector can have a \emph{non-vanishing
trace}.\vskip4pt

In \citep{Chauvineau-2003,Chauvineau-2007} one of the present authors
(BC) introduced a set of ideas aimed at establishing the ${\cal O}\left(1/\sqrt{\omega}\right)$
behavior for matter with non-vanishing EMT trace, such as in the action
\eqref{eq:action-generic}. This section builds upon those ideas,
working towards a comprehensive proof. Departing from the conventional
expansion in terms of $1/\omega$, we initiate our proof, as made
in \citep{Chauvineau-2003} for the scalar and the metric, with the
following expansions:
\begin{align}
\phi & =\bar{\phi}\,\Bigl(1+\frac{\stackrel{1}{\varphi}}{\sqrt{\omega}}+\frac{\stackrel{2}{\varphi}}{\omega}+\dots\Bigr)\label{eq:expension-phi}\\
g_{\mu\nu} & =\ \stackrel{0}{g}_{\mu\nu}+\text{\,}\frac{\stackrel{1}{g}_{\mu\nu}}{\sqrt{\omega}}+\text{\,}\frac{\stackrel{2}{g}_{\mu\nu}}{\omega}+\dots\label{eq:expansion-g}\\
\Psi & =\ \stackrel{0}{\Psi}+\,\frac{\stackrel{1}{\Psi}}{\sqrt{\omega}}+\,\frac{\stackrel{2}{\Psi}}{\omega}+\dots\label{eq:expansion-Psi}
\end{align}
which \emph{supersede} Eqs. \eqref{eq:series-phi} and \eqref{eq:series-g}.
The metric $\stackrel{0}{g}_{\mu\nu}$ shall be referred to as the
``background'' metric. The critical questions at hand are two-fold:
(i) whether the ``background'' metric $\stackrel{0}{g}_{\mu\nu}$
corresponds to the GR solution with only $\stackrel{0}{\Psi}$ as
the matter field, and (ii) whether $\stackrel{1}{\varphi}$ exhibits
variation. (Note: A constant value of $\stackrel{1}{\varphi}$ can
be formally removed by redefining $\bar{\phi}$.) A\emph{ non-constant}
field $\stackrel{1}{\varphi}$ would be sufficient to establish the
$\mathcal{O}\left(1/\sqrt{\omega}\right)$ hallmark.\vskip8pt

Plugging the Taylor expansions of $1/\sqrt{\omega}$ into the action
\eqref{eq:action-generic}
\begin{equation}
\mathcal{S}_{\text{generic}}=\ \stackrel{0}{\mathcal{S}}_{\text{generic}}+\mathcal{O}\left(\frac{1}{\sqrt{\omega}}\right)
\end{equation}
in which the leading term
\begin{align}
\stackrel{0}{\mathcal{S}}_{\text{generic}} & :=\int d^{4}x\sqrt{-\stackrel{0}{g}}\biggl[\bar{\phi}\Bigl(\stackrel{0}{\mathcal{R}}-\stackrel{0\ \ }{g^{\mu\nu}}\partial_{\mu}\stackrel{1}{\varphi}\partial_{\nu}\stackrel{1}{\varphi}\Bigr)\nonumber \\
 & \ \ \ \ \ \ \ \ \ \ \ \ \ \ \ \ \ \ \ \ \ \ \ -L^{(m)}\Bigl(\{\stackrel{0}{g}_{\alpha\beta},\bar{\phi}\};\stackrel{0}{\Psi}\Bigr)\biggr]\label{eq:action-0-generic}
\end{align}
Varying the action \eqref{eq:action-0-generic} results in the field
equations
\begin{align}
\stackrel{0}{\mathcal{R}}_{\mu\nu}\, & =\partial_{\mu}\stackrel{1}{\varphi}\partial_{\nu}\stackrel{1}{\varphi}+\bar{\phi}^{-1}\left[\,\stackrel{0}{T}_{\mu\nu}-\,\frac{1}{2}\stackrel{0}{g}_{\mu\nu}\,\stackrel{0}{T}\,\right]\\
\stackrel{0}{\square}\ \stackrel{1}{\varphi}\, & =0\label{eq:harmonic}
\end{align}
and a conservation law for the EMT (not shown here). The EMT of the
matter sector is defined as
\begin{equation}
\stackrel{0}{T}_{\mu\nu}:=\left.\frac{-2}{\sqrt{-g}}\frac{\delta\,\bigl(\sqrt{-g}\,L^{(m)}(\{g_{\alpha\beta},\phi\};\Psi)\bigr)}{\delta g^{\mu\nu}}\right|_{\left\{ \begin{array}{l}
g_{\mu\nu}=\,\stackrel{0}{g}_{\mu\nu}\\
\phi=\bar{\phi}\\
\Psi=\,\stackrel{0}{\Psi}
\end{array}\right.}\label{eq:EMT-def-1}
\end{equation}
which is nothing but the original EMT evaluated at $\stackrel{0}{g}_{\mu\nu}$,
$\bar{\phi}$, and $\stackrel{0}{\Psi}$. \emph{Note that the equation
for $\stackrel{1}{\varphi}$ is independent of the trace of the EMT.
}\footnote{We should emphasize that the sourceless ``harmonic'' equation, Eq.
\eqref{eq:harmonic}, holds in general without assuming stationarity
or spherical symmetry. Despite lacking a source, it allows for variations
in $\stackrel{1}{\varphi}$. For instance, in the context of a Minkowski
background where $\stackrel{0}{g}_{\mu\nu}=\text{diag}(-1,1,1,1)$,
the equation reduces to the d'Alembert equation, describing a \emph{free}
propagating wave for $\stackrel{1}{\varphi}$.}\emph{\vskip4pt}

In the absence of the BD field, $\stackrel{1}{\varphi}\ \equiv0$.
The background metric is that of GR coupled with the matter field
$\Psi$. In the presence of the BD field, the limiting configuration
deviates from GR (coupled with the matter field $\Psi$) due to the
contribution from $\stackrel{1}{\varphi}$. If $\stackrel{1}{\varphi}$
is non-constant then the $\mathcal{O}(1/\sqrt{\omega})$ signature
is established.\vskip4pt

For the static spherically symmetric setup, this statement can be
proved. The harmonic equation for $\stackrel{1}{\varphi}$ in the
harmonic radial coordinate yields a solution
\begin{equation}
\stackrel{1}{\varphi}\ \,=\,\sigma\,u\label{eq:varphi1}
\end{equation}
with $\sigma$ being a scalar charge. It should be commented that
in the harmonic radial coordinate, as per Eq. \eqref{eq:reciprocal-coord},
the spatial infinity corresponds to $u=0$. The divergence of $\stackrel{1}{\varphi}$
in Eq. \eqref{eq:varphi1} when $u$ goes to infinity means that the
solution is diverging at $r=0$. \emph{Consequently, the possibility
of the non--GR behavior, in this static case, is related to the presence
of a singularity at $r=0$.}\vskip4pt

The non-constant $\stackrel{1}{\varphi}$ thus adds a constant contribution
to the $uu-$equation, per
\begin{equation}
\stackrel{0}{\mathcal{R}}_{uu}\ =\,\sigma^{2}+\bar{\phi}^{-1}\left[\,\stackrel{0}{T}_{uu}-\,\frac{1}{2}\stackrel{0}{g}_{uu}\,\stackrel{0}{T}\,\right]
\end{equation}
Note that the $tt-$ and $\theta\theta-$equations are not affected.
Therefore, the functional form of the metric (in the radial harmonic
gauge) is unaffected. This is because the functional forms of $\gamma(u)$
and $\alpha(u)$ are fully determined by
\begin{align}
\stackrel{0}{\mathcal{R}}_{tt} & \ =\bar{\phi}^{-1}\left[\,\stackrel{0}{T}_{tt}-\,\frac{1}{2}\stackrel{0}{g}_{tt}\,\stackrel{0}{T}\,\right]\\
\stackrel{0}{\mathcal{R}}_{\theta\theta} & \ =\bar{\phi}^{-1}\left[\,\stackrel{0}{T}_{\theta\theta}-\,\frac{1}{2}\stackrel{0}{g}_{\theta\theta}\,\stackrel{0}{T}\,\right]
\end{align}
The $uu-$equation acts as a ``constraint''. The relation between
the parameters in $A(u)$ and $C(u)$ is altered due to the presence
of $\sigma^{2}$ in the new ``constraint''. The limiting configuration
$\stackrel{0}{g}_{\mu\nu}$ thus deviates from its usual one without
the BD field.\vskip4pt

This completes our proof. Note: The proof presented herein does not
depend on the trace of the matter source $\Psi$. Moreover, the signature
of $1/\sqrt{\omega}$ is self evident via the existence of $\stackrel{1}{\varphi}$.

\section{\label{sec:gamma}The Robertson parameters}

In this section, we derive the Robertson (or Eddington-Robertson-Schiff)
parameters, $\beta$ and $\gamma$, which are physical, i.e. \emph{measurable},
quantities, from our exact BDKG solution.\vskip4pt

For GR, it is known that $\beta_{\,\text{GR}}=\gamma_{\,\text{GR}}=1$.
The Robertson expansion in isotropic coordinates is, using the relativistic
units \citep{Weinberg}:
\begin{align}
ds^{2} & =-\left(1-2\,\frac{M}{\rho}+2\beta\,\frac{M^{2}}{\rho^{2}}+\dots\right)dt^{2}\nonumber \\
 & \ \ \ \,+\left(1+2\gamma\,\frac{M}{\rho}+\dots\right)\left(d\rho^{2}+\rho^{2}d\Omega^{2}\right)\label{eq:Weinberg}
\end{align}
in which $\beta$ and $\gamma$ are the Robertson parameters. The
metric of Eq. \eqref{eq:app-1} in Appendix \ref{sec:Isotropic} yields
\begin{align}
\phi_{0}\,ds^{2} & =-\biggl(1-\frac{4K\left[(a+\sigma)L+(a-\sigma)\right]}{\sqrt{\chi}\left(1+L\right)\rho}\nonumber \\
 & +\frac{8K^{2}\left[(a+\sigma)^{2}L^{2}+2a^{2}L+(a-\sigma)^{2}\right]}{\chi\,(1+L)^{2}\rho^{2}}+\dots\biggr)\nonumber \\
 & \times\left(1+L\right)^{2}dt^{2}\nonumber \\
 & +\biggl(1+\frac{4K\left[(a-\sigma)L+(a+\sigma)\right]}{\sqrt{\chi}\left(1+L\right)\rho}+\dots\biggr)\nonumber \\
 & \times\left(1+L\right)^{2}\left(d\rho^{2}+\rho^{2}d\Omega^{2}\right)\label{eq:expansion}
\end{align}
where $L:=\frac{\Psi_{*}^{2}\phi_{0}}{8(\chi-a^{2})}$. Upon absorbing
$(1+L)$ into $dt$, $d\rho$ and $\rho$, and $\phi_{0}$ into $ds^{2}$
in Eq. \eqref{eq:expansion} and comparing it with Eq. \eqref{eq:Weinberg},
we have
\begin{align}
M & =\frac{2K}{\sqrt{\chi}}\left[(a+\sigma)L+(a-\sigma)\right]\\
\beta & =1+\frac{2\sigma^{2}L}{\left[(a+\sigma)L+(a-\sigma)\right]^{2}}\\
\gamma & =1-\frac{2\sigma\left(L-1\right)}{(a+\sigma)L+(a-\sigma)}
\end{align}
Employing the relation \eqref{eq:app-2}, the Robertson parameters
become
\begin{align}
\beta & =1+\frac{\Psi_{*}^{2}\phi_{0}/(4a^{2})}{\Bigl[\Bigl(\frac{\Psi_{*}^{2}\phi_{0}}{8(\chi-a^{2})}+1\Bigr)\sqrt{2\omega+3}+\Bigl(\frac{\Psi_{*}^{2}\phi_{0}}{8(\chi-a^{2})}-1\Bigr)\frac{\sqrt{\chi-a^{2}}}{a}\Bigr]^{2}}\\
\gamma & =1-\frac{2\left(\frac{\Psi_{*}^{2}\phi_{0}}{8\left(\chi-a^{2}\right)}-1\right)\frac{\sqrt{\chi-a^{2}}}{a}}{\Bigl(\frac{\Psi_{*}^{2}\phi_{0}}{8(\chi-a^{2})}+1\Bigr)\sqrt{2\omega+3}+\Bigl(\frac{\Psi_{*}^{2}\phi_{0}}{8(\chi-a^{2})}-1\Bigr)\frac{\sqrt{\chi-a^{2}}}{a}}
\end{align}
These formulae evidently recover $\beta_{\,\text{GR}}=\gamma_{\,\text{GR}}=1$
in the infinite $\omega$ limit, and exhibit behavior on order of
$\mathcal{O}\left(1/\sqrt{\omega}\right)$ for a fixed set of $a$
and $\chi$. As physical quantities, $\beta$ measures the degree
of nonlinearity in superposition of gravity, while $\gamma$ specifies
the amount of spatial curvature produced by the mass source \citep{Will1,Will2}.
Therefore, the $\mathcal{O}\left(1/\sqrt{\omega}\right)$ signature
in $\beta$ and $\gamma$ carry meaningful physical implications in
BD configurations involving trace-carrying KG matter fields.

\section{\label{sec:Physical-implications}Physical implications}

Besides the $\mathcal{O}\left(1/\sqrt{\omega}\right)$ imprints in
the Roberson parameters $\beta$ and $\gamma$ as presented in the
preceding section, let us further stress that the issue tackled in
this paper is not just a question of formal mathematics, but that
genuine potentially new physical features are at stake, in both the
EMT trace $T=0$ and $T\neq0$ cases. Indeed, it results from an $\mathcal{O}\left(1/\sqrt{\omega}\right)$
behavior of the BD scalar field that the large $\omega$ Brans-Dicke
vacuum spherical solution (i.e. Brans class I if $\omega>-3/2$) does
\emph{not} converge to the GR vacuum spherical solution (i.e. Schwarzschild),
but to the GR spherical solution filled by a massless scalar (i.e.
the FJNW solution). Especially in the strongly scalarized case, the
FJNW solution exhibits drastically specific features with respect
to Schwarzschild. Indeed, for a given gravitational mass of the field,
circular orbits exist for any areal radius (while the areal radius
is lower bounded in Schwarzschild), with a diverging orbital frequency
measured by a remote observer (while this frequency is upper bounded
in Schwarzschild), despite the Einstein gravitational red-shifting
\citep{Chauvineau-2017}. Since the associated gravitational wave
(GW) frequency, in the Extreme Mass Ration Inspiral (EMRI) scheme,
is twice the orbital one, the potential impact of the effect in the
new GW astronomy is obvious. Still in the strongly scalarized case,
another qualitative specific features with respect to Schwarzschild
concerns the light bending, which experiments a maxima when the impact
parameter decreases. This results in the presence of a caustic surface,
which yields an infinite amplification of the apparent luminosity
of a pointlike star crossing this caustics \citep{Chauvineau-2022}.
Thence, the case of an orbiting star on a properly inclined orbit
results in a sequence of luminosity peaks, an effect which does not
exist in the Schwarzschild spacetime. In the $T\neq0$ case, the age
of the Universe is not constrained to a fixed value by the present
Hubble parameter $H_{0}$ in BD dust filled flat Robertson-Walker
cosmology, even in the $\omega\rightarrow\infty$ limit, while it is fixed
to $2/(3H_{0})$ in GR dust filled flat Robertson-Walker cosmology
\citep{Chauvineau-2007}. We plan to visit this issue in the future.

\section{\label{sec:Conclusion}Conclusion and outlook}

The infinite $\omega$ limit in BD gravity is a nuanced matter, encapsulated
by the anomalous ${\cal O}\left(1/\sqrt{\omega}\right)$ behavior
identified in BD vacuum and electrovacuum since the 1990s \citep{Romero-1993-a,Romero-1993-b,Romero-1993-c,Romero-1998,Romero-2021,Faraoni-1998,Faraoni-1999,Faraoni-2019,Bhadra-2001,Bhadra-2002,Bhadra-2005,Fabris-2019,Quiros-1999,Banerjee-1997,Faraoni-2018}.
Gaining a clear understanding of this phenomenon could bear significant
implications for scalar-tensor theories at large.\vskip4pt

Textbook treatments typically\emph{ assume} a diminishing behavior
of the BD scalar field $\phi$ in an ${\cal O}\left(1/\omega\right)$
fashion \citep{Weinberg}. Due to this presumed requirement, in the
infinite $\omega$ limit, the ``kinetic'' term $\omega\left(\nabla\phi/\phi\right)^{2}$
in the BD action would be suppressed compared with the Einstein-Hilbert
term ${\cal R}$, resulting in a recovery to GR with the same matter
content. However, from the variational principle standpoint, this
scenario is not \emph{a priori} guaranteed. As the BD action is extremized,
the ``kinetic'' term and the Einstein-Hilbert term can be in competition,
leading to an alternative scenario where the ``kinetic'' term retains
a finite and comparable contribution. This latter outcome necessarily
compels the BD field to behave as
\begin{equation}
\phi\simeq\bar{\phi}\,\biggl[1+\frac{\stackrel{1}{\varphi}}{\sqrt{\omega}}+{\cal O}\left(\frac{1}{\omega}\right)\biggr]
\end{equation}
with the sub-leading field $\stackrel{1}{\varphi}$ being \emph{non-constant},
and the ``kinetic'' term surviving the infinite $\omega$ limit, per
$\omega\left(\nabla\phi/\phi\right)^{2}\simeq\bigl(\ \nabla\stackrel{1}{\varphi}\:\bigr)^{2}$.\vskip4pt

The proof presented in Section \ref{sec:Formal-proof} solidifies
this behavior, elucidating its origin and demonstrating that it persists
\emph{irrespective} of the EMT trace of the matter source.\vskip4pt

The ``remnant'' field $\stackrel{1}{\varphi}$ emerges as a free massless
scalar field, leaving its footprints in the limiting spacetime configuration
as $\omega\rightarrow\infty$. In the broader context of Bergmann-Wagoner
theory, the possibility of such a ``remnant'' field emerging and,
if so, its role in the Damour-Nordtvedt attractor mechanism would
deserve further examination.\vskip8pt

In BD gravity, the presence of a ``non-analytical'' square-root-of-$\omega$
term may initially seem surprising. However, such a term, viz. $\sqrt{2\omega+3}$,
arises during the transition of the action from the Jordan frame to
the Einstein frame via a Weyl mapping. The $\sqrt{\omega}$ signature
is inherent in the gravitational sector and is unrelated to the matter
sector. This is the key reason underlying its prevalence \emph{regardless}
of the EMT trace of matter. We explained this important point in Section
\ref{sec:Recap}\vskip4pt

Along the same line of reasoning, for the generic Bergmann-Wagoner
action \eqref{eq:BW-action}, the shift to the Einstein frame, viz.
$\sqrt{-\tilde{g}}\left[\tilde{{\cal R}}-\frac{1}{2}\left(\tilde{\nabla}\varphi\right)^{2}...\right]$,
entails a Weyl mapping given by
\begin{equation}
\varphi=\varphi_{0}-\frac{1}{2}\int_{\phi_{0}}^{\phi}\frac{d\phi}{\phi}\sqrt{2\omega(\phi)+3}
\end{equation}
which is also exclusive to the gravitational sector. Hence, for a
given function $\omega(\phi)$, any properties that are result from
the $\sqrt{2\omega(\phi)+3}$ term should be universal, independent
of the EMT trace of matter.\vskip4pt

The proof presented in Section \ref{sec:Formal-proof} is general.
We further demonstrated its validity through a specific case---the
BD action coupled with a massless Klein Gordon scalar field (the EMT
of which has non-vanishing trace). In Section \ref{sec:BDKG-action},
we expanded upon Bronnikov's approach \citep{Bronnikov-1973} by utilizing
the Einstein frame and the harmonic radial coordinate. Despite the
complication of a dilatonic coupling between the Einstein-frame BD
field and the Klein-Gordon field, the system of field equations prove
to be soluble. We obtained a static spherically symmetric solution
for this coupled action, which can be useful for future studies of
BD theory sourced by trace-carrying matter. For the purpose at hand,
the characteristic signature of $1/\sqrt{\omega}$ is distinctly evident
in the final solution.\vskip4pt

Possible extensions of our solution to include Maxwell EM field are
deferred for future research. Related studies in GR, coupled with
the Maxwell field and a dilaton field, have been reported in \citep{Clement-2009}.\vskip12pt

\textbf{\emph{Outlook---}}We discussed the physical implications
of the ${\cal O}\left(1/\sqrt{\omega}\right)$ anomaly in Section
\ref{sec:Physical-implications}. There are other possibilities that
warrant further investigation. In particular, the Robertson parameter
$\gamma$ of BD gravity has been derived via the parametrized post-Newtonian
(PPN) approximation to be $\gamma_{PPN}=\frac{\omega+1}{\omega+2}$,
which lacks the ${\cal O}\left(1/\sqrt{\omega}\right)$ signature.
This should prompt a scrutiny of the validity of the PPN result, including
the underlying assumptions in its derivation. Notably, a recent study
has challenged the weak-field assumption inherent in the PPN approximation
\citep{Faraoni-2020}. We have also delved into this matter in two recent reports \citep{Nguyen-compact-star-1,Nguyen-compact-star-2}.
\begin{acknowledgments}
We are grateful to the anonymous referees for their insightful feedbacks
which helped improve the quality of this paper. H.K.N. thank Valerio
Faraoni for helpfully pointing out a number of crucial references,
and Mustapha Azreg-A\"inou, Tao Zhu, Luis A. Correa-Borbonet for
stimulating discussions.
\end{acknowledgments}

\begin{center}
------------$\infty$------------
\par\end{center}

\appendix

\section{\label{sec:Field-equations}Field equations of Einstein-Klein-Gordon-dilation
(EKGD) action}

This Appendix derives the field equations for the EKGD action \eqref{eq:EKGD-action}.
Upon variation
\begin{equation}
\delta\left(\sqrt{-\tilde{g}}\mathcal{\tilde{R}}\right)=\sqrt{-\tilde{g}}\left(\frac{1}{2}\tilde{g}^{\mu\nu}\mathcal{\tilde{R}}-\tilde{\mathcal{R}}^{\mu\nu}\right)\delta\tilde{g}_{\mu\nu}+\text{total deriv.}
\end{equation}
For other terms, denoting the symmetric rank-2 tensor $A_{\mu\nu}:=\left(\omega+3/2\right)\partial_{\mu}\theta\partial_{\nu}\theta+e^{-\theta}\partial_{\mu}\Psi\partial_{\nu}\Psi$
and $A:=\tilde{g}^{\mu\nu}A_{\mu\nu}$ the trace of $A_{\mu\nu}$,
we have
\begin{align}
 & \delta\left(\sqrt{-\tilde{g}}\tilde{g}^{\mu\nu}A_{\mu\nu}\right)\nonumber \\
 & =\sqrt{-\tilde{g}}\left(\frac{1}{2}\tilde{g}^{\mu\nu}A-A^{\mu\nu}\right)\delta\tilde{g}_{\mu\nu}+\sqrt{-\tilde{g}}\tilde{g}^{\mu\nu}\delta A_{\mu\nu}
\end{align}
Regarding the tensor $A_{\mu\nu}$:
\begin{equation}
\delta A_{\mu\nu}=(\omega+3/2)\,\delta\left(\partial_{\mu}\theta\partial_{\nu}\theta\right)+\delta\left(e^{-\theta}\partial_{\mu}\Psi\partial_{\nu}\Psi\right)
\end{equation}
the first term is
\begin{align}
\delta\left(\partial_{\mu}\theta\partial_{\nu}\theta\right) & =\left[\tilde{\nabla}_{\mu}\left(\delta\theta\partial_{\nu}\theta\right)+\tilde{\nabla}_{\nu}\left(\delta\theta\partial_{\mu}\theta\right)\right]-2(\tilde{\nabla}_{\mu}\tilde{\nabla}_{\nu}\theta)\,\delta\theta
\end{align}
This results in, up to a total derivative
\begin{align}
\sqrt{-\tilde{g}}\tilde{g}^{\mu\nu}\delta\left(\partial_{\mu}\theta\partial_{\nu}\theta\right) & =-2\sqrt{-\tilde{g}}\tilde{\square}\,\theta\,\delta\theta\label{eq:s-1}
\end{align}
and
\begin{align}
\delta\left(e^{-\theta}\partial_{\mu}\Psi\partial_{\nu}\Psi\right) & =e^{-\theta}\delta\left(\partial_{\mu}\Psi\partial_{\nu}\Psi\right)-\delta\theta e^{-\theta}\partial_{\mu}\Psi\partial_{\nu}\Psi\,.
\end{align}
Similarly for Eq. \eqref{eq:s-1}, up to a total derivative
\begin{align}
 & \sqrt{-\tilde{g}}\tilde{g}^{\mu\nu}e^{-\theta}\delta\left(\partial_{\mu}\Psi\partial_{\nu}\Psi\right)\nonumber \\
 & =2e^{-\theta}\partial_{\mu}\theta\sqrt{-\tilde{g}}\delta\Psi\tilde{\nabla}^{\mu}\Psi-2\sqrt{-\tilde{g}}e^{-\theta}\tilde{\square}\,\Psi\,\delta\Psi
\end{align}
leading to 
\begin{align}
 & \sqrt{-\tilde{g}}\tilde{g}^{\mu\nu}\delta\left(e^{-\theta}\partial_{\mu}\Psi\partial_{\nu}\Psi\right)\nonumber \\
 & =\sqrt{-\tilde{g}}e^{-\theta}\left[2(\partial_{\mu}\theta\tilde{\nabla}^{\mu}\Psi-\tilde{\square}\,\Psi)\delta\Psi-(\tilde{\nabla}\Psi)^{2}\,\delta\theta\right]
\end{align}
Combining all terms, the variation of the action \eqref{eq:EKGD-action}
is
\begin{align}
\sqrt{-\tilde{g}}\biggl[\left(\frac{1}{2}\tilde{g}^{\mu\nu}\mathcal{\tilde{R}}-\tilde{\mathcal{R}}^{\mu\nu}\right)-\left(\frac{1}{2}\tilde{g}^{\mu\nu}A-A^{\mu\nu}\right)\biggr]\,\delta\tilde{g}_{\mu\nu}\nonumber \\
+\Bigl[(2\omega+3)\sqrt{-\tilde{g}}\tilde{\square}\,\theta+\sqrt{-\tilde{g}}e^{-\theta}(\tilde{\nabla}\Psi)^{2}\Bigr]\,\delta\theta\label{eq:variations}\\
+2\sqrt{-\tilde{g}}e^{-\theta}(-\partial_{\mu}\theta\tilde{\nabla}^{\mu}\Psi+\tilde{\square}\,\Psi)\,\delta\Psi\nonumber 
\end{align}
Imposing the extremality on \eqref{eq:variations}, regarding $\delta\tilde{g}_{\mu\nu}$,
we obtain
\begin{align}
\mathcal{\tilde{R}}^{\mu\nu}-\frac{1}{2}\tilde{g}^{\mu\nu}\mathcal{\tilde{R}} & =A^{\mu\nu}-\frac{1}{2}\tilde{g}^{\mu\nu}A
\end{align}
which, upon taking the trace, $\mathcal{\tilde{R}}=A$, results in
$\tilde{\mathcal{R}}_{\mu\nu}=A_{\mu\nu}$, or
\begin{equation}
\mathcal{\tilde{R}}_{\mu\nu}=\left(\omega+3/2\right)\partial_{\mu}\theta\partial_{\nu}\theta+e^{-\theta}\partial_{\mu}\Psi\partial_{\nu}\Psi\label{eq:EF-eqn-R}
\end{equation}
Regarding $\delta\theta$, Eq. \eqref{eq:variations} yields
\begin{equation}
(2\omega+3)\,\tilde{\square}\,\theta=-e^{-\theta}(\tilde{\nabla}\Psi)^{2}\label{eq:EF-eqn-theta}
\end{equation}
Regarding $\delta\Psi$, Eq. \eqref{eq:variations} yields
\begin{equation}
\tilde{\square}\,\Psi=\tilde{\nabla}^{\mu}\Psi\tilde{\nabla}_{\mu}\theta\label{eq:EF-eqn-Psi}
\end{equation}
The last two equations can also be neatly cast as
\begin{align}
(2\omega+3)\,e^{-\theta}\,\tilde{\square}\,\theta & =-(e^{-\theta}\tilde{\nabla}^{\mu}\Psi)\,(e^{-\theta}\tilde{\nabla}_{\mu}\Psi)\label{eq:e-1}\\
\tilde{\nabla}^{\mu}\left(e^{-\theta}\tilde{\nabla}_{\mu}\Psi\right) & =e^{-\theta}\left(-\tilde{\nabla}^{\mu}\theta\tilde{\nabla}_{\mu}\Psi+\tilde{\square}\Psi\right)=0\label{eq:e-2}
\end{align}

When a non-constant Klein-Gordon (KG) field $\Psi$ is present, Eq.$\,$\eqref{eq:EF-eqn-theta}
requires that the `dilaton' field $\theta$ must also be non-constant.

\section{\label{sec:Isotropic}$\ $The new exact solution of BDKG action
in isotropic coordinates}

For many future purposes, it can be useful to express the exact solution
\eqref{eq:BDKG-solution}--\eqref{eq:KG-scalar-solution} in isotropic
form. Since one is finally interested in the large $\omega$ limit,
let us specify to the $\omega>-3/2$ case (which, besides, is the
condition for the BD theory to be Ostrogradski stable \citep{Woodard-2007}).\vskip4pt

This is achieved by the following radial coordinate transformation
\begin{equation}
\left\{ \begin{array}{l}
\rho=K\frac{e^{\sqrt{\chi}u}+1}{e^{\sqrt{\chi}u}-1}\text{ \ \ \ensuremath{\Longleftrightarrow}\ \ }e^{\sqrt{\chi}u}=\frac{\rho+K}{\rho-K}\\
K>0\\
\rho\in\left[K,+\infty\right]
\end{array}\right.
\end{equation}
Let us remark that $\rho=\infty$ for $u=0$. The solution \eqref{eq:BDKG-solution}--\eqref{eq:KG-scalar-solution}
then achieves the form, rescaling $dt$ and $ds$ by a $2K/\sqrt{\chi}$
factor
\begin{equation}
\left\{ \begin{array}{ll}
ds^{2} & =\frac{1}{\phi(\rho)}\biggl[-\Bigl(\frac{\rho-K}{\rho+K}\Bigr)^{\frac{2a}{\sqrt{\chi}}}dt^{2}\\
 & +\Bigl(\frac{\rho^{2}-K^{2}}{\rho^{2}}\Bigr)^{2}\Bigl(\frac{\rho-K}{\rho+K}\Bigr)^{-\frac{2a}{\sqrt{\chi}}}\bigl(d\rho^{2}+\rho^{2}d\Omega^{2}\bigr)\biggr]\\
\phi(\rho) & =\phi_{0}\frac{\left(\frac{\rho-K}{\rho+K}\right)^{\frac{2\sigma}{\sqrt{\chi}}}}{\biggl[1+\frac{\Psi_{\ast}^{2}\phi_{0}}{8(\chi-a^{2})}\left(\frac{\rho-K}{\rho+K}\right)^{\frac{2\sigma}{\sqrt{\chi}}}\biggr]^{2}}\\
\Psi(\rho) & =-\frac{\Psi_{\ast}\phi_{0}}{2\sigma}\frac{\left(\frac{\rho-K}{\rho+K}\right)^{\frac{2\sigma}{\sqrt{\chi}}}}{1+\frac{\Psi_{\ast}^{2}\phi_{0}}{8(\chi-a^{2})}\left(\frac{\rho-K}{\rho+K}\right)^{\frac{2\sigma}{\sqrt{\chi}}}}
\end{array}\right.\label{eq:app-1}
\end{equation}
where
\begin{equation}
\sigma\equiv\sqrt{\frac{\chi-a^{2}}{2\omega+3}}\label{eq:app-2}
\end{equation}
Since a global constant factor in the metric is not of significance
(it just corresponds to a proper time's change of units), one can
also rewrite the metric as
\begin{align}
ds^{2} & =\frac{\phi_{1}}{\phi(\rho)}\biggl[-\Bigl(\frac{\rho-K}{\rho+K}\Bigr)^{\frac{2a}{\sqrt{\chi}}}dt^{2}\nonumber \\
 & +\Bigl(\frac{\rho^{2}-K^{2}}{\rho^{2}}\Bigr)^{2}\Bigl(\frac{\rho-K}{\rho+K}\Bigr)^{-\frac{2a}{\sqrt{\chi}}}\bigl(d\rho^{2}+\rho^{2}d\Omega^{2}\bigr)\biggr]
\end{align}
with
\begin{equation}
\phi_{1}=\phi\left(\rho=\infty\right)=\phi_{0}\,\left(1+\frac{\Psi_{\ast}^{2}\phi_{0}}{8(\chi-a^{2})}\right)^{-2}
\end{equation}
for the metric to achieve the standard form ($ds^{2}=-dt^{2}+d\rho^{2}+\rho^{2}d\Omega^{2}$)
at spatial infinity.

\end{document}